\documentstyle[12pt]{ioplppt}
\newcommand{\D}{\mbox{d}}
\newcommand{\HRG}{H^{\mbox{\scriptsize RG}}}
\newcommand{\HRGN}[1]{H^{\mbox{\scriptsize RG{#1}}}}
\newcommand{\MN}[1]{M^{\mbox{\scriptsize {#1}}}}
\newcommand{\al}[1]{\alpha_{#1}}

\newcommand{\fracp}[2]{\frac{\partial {#1}}{\partial {#2}}}
\newcommand{\B}[1]{B_{#1}}
\newcommand{\C}[1]{C_{#1}}
\newcommand{\F}[2]{\frac{#1}{#2}}
\newcommand{\NO}{\nonumber\\}
\newcommand{\U}[2]{u_{#1}^{(#2)}}
\newcommand{\V}[2]{v_{#1}^{(#2)}}

\newcommand{\uz}{u^{(0)}}
\newcommand{\QZ}{\vec{q}_{0}}
\newcommand{\DT}{\D t}
\begin{document}
%%%% DPNU number %%%%
%\begin{flushright}
% DPNU-98-29 \\
% September 1998
%\end{flushright}
%%%%%%%%%%%%%%%%%%%%%
% IOP Journal of Physics
\jl{1}
%%%%%%%%%%%%%%%%%%%%%
\title{Canonical structure of renormalization group equations 
and separability of Hamiltonian systems}[Canonical structure of renormalization
group equations]

\author{Yoshiyuki Y. YAMAGUCHI\S
\ftnote{1}{e-mail: yamaguchi@allegro.phys.nagoya-u.ac.jp}
  and Yasusada NAMBU\P
\ftnote{2}{e-mail: nambu@allegro.phys.nagoya-u.ac.jp}}

\address{\S\ The general research organization of science and
  engineering, Ritsumeikan University,  
  Kusatsu, Shiga 525-8577, Japan}
\address{\P\  Department of Physics, 
  Nagoya  University, Nagoya 464-8602, Japan}
%%%
%%%
%%%
\newpage
%%%
\begin{abstract}
We investigate perturbed Hamiltonian systems 
with two degrees of freedom by renormalization group method,
which derives a reduced equation called 
renormalization group equation (RGE) by handling secular terms.
We found that RGE is not always a Hamiltonian system. 
The necessary and sufficient condition 
that RGE becomes a Hamiltonian system up to the second leading order
of a small parameter is that the original system is
separable by Cartesian coordinates. 
 Moreover, RGE is integrable when it is
a Hamiltonian system. These statements are partial generalizations 
of our previous paper.
\end{abstract}
\maketitle

%%%%%%%%%%%%%%%%%%%%%%%%%%%%%%%%%%%%%%%%%%%%%%%%%%%%%%%%%%%%%%%
%% Section 1
%%%%%%%%%%%%%%%%%%%%%%%%%%%%%%%%%%%%%%%%%%%%%%%%%%%%%%%%%%%%%%%
\section{Introduction}
\label{sec:introduction}

To understand the nature of non-linearity in dynamical systems, it 
is indispensable to investigate temporal 
evolutions of the system. Particularly, in Hamiltonian systems,
slow relaxation of $1/f$ spectra is observed
in systems with two \cite{karney-83,chirikov-84}
and many \cite{baba-97,yamaguchi-97} degrees of freedom,
and  understanding its cause 
is an important subject of Hamiltonian dynamical systems.  
The slow relaxation is connected with self-similar hierarchical 
structure of phase space,
which is observed in a 2-dimensional mapping \cite{ruffo-96}.
The structure is also supposed in systems with many degrees of freedom
with the aid of Kolmogorov-Arnold-Moser theorem \cite{kam}
and Poincar\'e-Birkhoff theorem \cite{lichtenberg-92}.
By assuming self-similar hierarchical structure
and transition probabilities between levels of the hierarchy,
some abstract models \cite{meiss-86,aizawa-84,yamaguchi-98a}
are proposed to understand the slow relaxation.
The abstract models catch universal properties of Hamiltonian 
systems,
but they are not available to investigate
details of individual systems owing to their assumptions.
For instance, Poincar\'e recurrences $P(t)$
(integrated probability to return into a given region 
after a time larger than $t$)
decays with $t$ as $P(t)\sim 1/t^{p}$,
and the exponent $p$ takes different values 
in different 2-dimensional mappings \cite{chirikov-84}.
One approach to grasp such individual properties
is to numerically calculate equations of motion,
and another approach is to analytically construct approximate
solutions. 
In this paper we focus on the latter approach
in nearly integrable systems with perturbation techniques.

Naive perturbation often yields secular terms
due to resonances 
and they invalidate perturbative expansion.
This problem of secular terms is dealt with
singular perturbation techniques \cite{hinch-91},
e.g. averaging methods, multiple scale methods
and matched asymptotic expansions.
However, these techniques are not easy to use
because we must select a suitable assumption about the 
structure of perturbation series.
 Recently, perturbative renormalization group method 
\cite{chen-96,kunihiro-95} is proposed
as a tool for global asymptotic analysis of the solution of 
differential equations.
 It unifies the methods listed
above, and can treat many systems irrespective of their features.
Renormalization group method omits fast motion 
and constructs reduced equations of motion from secular terms.
The reduced equation is called 
renormalization group equation (RGE),
which is geometrically regarded as an envelope equation.

Although RGE gives approximate but global solutions to original
systems,  we have one question:
whether does RGE keep properties of original systems?
We miss characteristic behaviors of orbits in phase space
if  RGE becomes a dissipative system which often has attractors and 
orbits exponentially converge to them. 
In contrast, Hamiltonian systems never have attractors
and orbits eternally wander.

In the previous paper \cite{yamaguchi-98b}, 
we treated a Hamiltonian system with a
 homogeneous cubic or quartic potential function 
 added as perturbation, and showed that if
the RGE becomes a Hamiltonian system 
 then the original Hamiltonian system is integrable. 
By investigating properties of RGE, we can justify integrability
 of the original Hamiltonian system. 
The present paper aim to extend the result  for general perturbation
function $V(q_1, q_2)$. Moreover, we mention integrability and an
integral of RGE.

The plan of this paper is as follows.
In section \ref{sec:RG-method} we review renormalization group method
by using a simple example.
The main theorem is shown in section \ref{sec:main-theorem},
and it is proved in sections \ref{sec:AtoB}, \ref{sec:BtoA}
and \ref{sec:integral}.
Section \ref{sec:summary} is devoted to summary and discussions.

%%%%%%%%%%%%%%%%%%%%%%%%%%%%%%%%%%%%%%%%%%%%%%%%%%%%%%%%%%%%%%%
%% Section 2
%%%%%%%%%%%%%%%%%%%%%%%%%%%%%%%%%%%%%%%%%%%%%%%%%%%%%%%%%%%%%%%
\section{Renormalization Group Method}
\label{sec:RG-method}

Here we review renormalization group method by using a simple system.
Let us consider the system
%%%%
\begin{equation}
        \label{system-ex}
  H(q,p) = H_0(q,p) + \epsilon V(q), 
\end{equation}
%%%%
where
%%%%
\begin{equation}
  H_0(q,p) =\frac{1}{2}(p^2 + q^2),~~~~  V(q) = \frac{1}{2} q^2,
\end{equation}
%%%%
and $\epsilon$ is a small parameter.
The equation of motion is 
%%%%
\begin{equation}
  \label{eq-V2}
  \frac{\D^2 q}{\D t^2} + q = -\epsilon q,
\end{equation}
%%%%
and the exact solution is
%%%%
\begin{equation}
  \label{exact-sol}
  q = B_0 \cos(\sqrt{1+\epsilon}\ t) + C_0 \sin(\sqrt{1+\epsilon}\
  t),
\end{equation}
%%%%
where $B_0$ and $C_0$ are constants of integration 
and determined by initial condition at the initial time $t=t_0$.

We perturbatively solve equation (\ref{eq-V2}) by expanding $q$ 
as a series of positive powers of $\epsilon$:
%%%%
\begin{equation}
  \label{expand}
  q = q_0 + \epsilon q_1 + \epsilon^2 q_2 + \cdots.
\end{equation}
%%%%
This naive expansion gives
%%%%
\begin{equation}
  \label{naive-sol}
\eqalign{
\fl
  q(t;t_0,B_0,C_0)
        = B_0 \cos t + C_0 \sin t + \frac{\epsilon}{2}(t-t_0)
            \left( C_0 \cos t - B_0 \sin t \right) \\
   \lo- \frac{\epsilon^2}{8} \left[
      (t-t_0) \left( C_0 \cos t - B_0 \sin t \right)
      + (t-t_0)^2 \left( B_0 \cos t + C_0 \sin t \right) \right] \\
    \lo+ O(\epsilon^3).
}
\end{equation}
%%%%
We ignored other homogeneous parts of $q_j\ (j\geq 1)$,
which are kernel of the linear operator $L\equiv \D^2/\D t^2 + 1$,
since they can be included in $q_0$.
% Non-determinicity of homogeneous part is a crucual problem
% of this renormalization group method.
% We will clarify this problem in the future article \ref{??}.
This naive expansion breaks when $\epsilon (t-t_0) \geq 1$
because amplitude of $\epsilon q_1$ exceeds one of $q_0$.

To remove the secular terms,
we regard $B_0$ and $C_0$ as functions of the initial time $t_0$.
The temporal evolutions of $B(t_0)$ and $C(t_0)$ are determined 
by the following equation \cite{chen-96,kunihiro-95}:
%%%%
\begin{equation}
  \label{renormalization}
  \left.\frac{\partial q}{\partial t_0}\right|_{t_0=t} = 0,
  \quad\mbox{for} ~~{}^\forall t.
\end{equation}
%%%%
From the equation (\ref{renormalization}), we obtain the RGE
%%%%
\begin{equation}
\label{renormal-ex} 
\eqalign{ 
  \frac{\D B(t)}{\D t} &= \left(\frac{\epsilon}{2} -
    \frac{\epsilon^2}{8}\right) C(t) + O(\epsilon^3),\\ 
  \frac{\D C(t)}{\D t} &= -\left(\frac{\epsilon}{2} -
    \frac{\epsilon^2}{8}\right) B(t) + O(\epsilon^3),
  }
\end{equation}
%%%%
where $B(t_0)=B_0$ and $C(t_0)=C_0$.
The renormalized solution $q^{\mbox{\scriptsize RG}}$ is
%%%%
\begin{eqnarray}
  q^{\mbox{\scriptsize RG}} &=&
        q(t;t,B(t),C(t)) \NO
   &=& B_0\cos\left( 
    \left(1+\frac{\epsilon}{2}-\frac{\epsilon^2}{8}\right)\ t \right) 
  + C_0\sin\left( 
    \left(1+\frac{\epsilon}{2}-\frac{\epsilon^2}{8}\right)\ t \right)
         + O(\epsilon^3), 
\end{eqnarray}
%%%%
and this solution is the same as the exact solution
(\ref{exact-sol}) up to $O(\epsilon^2)$. 

Let us check whether the reduced equation (\ref{renormal-ex})
reflects properties of the original system (\ref{system-ex}),
which are (i) symplectic properties, (ii) integrability
and (iii) the form of an integral.
In equation (\ref{renormal-ex}), $B$ and $C$ are canonical conjugate
variables and the equation is governed by the Hamiltonian
%%%%
\begin{equation}
  \HRG = \frac{1}{2} \left( \frac{\epsilon}{2}-\frac{\epsilon^2}{8} 
    \right) (B^2+C^2),
\end{equation}
%%%%
and canonical equations
%%%%
\begin{equation}
\eqalign{
  \frac{\D B(t)}{\D t} &=  \fracp{\HRG}{C}, \\
  \frac{\D C(t)}{\D t} &=  - \fracp{\HRG}{B}.
}
\end{equation}
%%%%
Equation (\ref{renormal-ex}) is therefore integrable 
and an integral of the system is $B^2+C^2$ which 
 corresponds to $H_0$.
As shown in section \ref{sec:BtoA}
these statements hold in any Hamiltonian systems
with one degree of freedom 
whose integrable part is a harmonic oscillator.
What happens in systems with two degrees of freedom
where chaos may appear ?
This is the main question of this article 
and we show a part of the answer in the next section.

%%%%%%%%%%%%%%%%%%%%%%%%%%%%%%%%%%%%%%%%%%%%%%%%%%%%%%%%%%%%%%%
%% Section 3
%%%%%%%%%%%%%%%%%%%%%%%%%%%%%%%%%%%%%%%%%%%%%%%%%%%%%%%%%%%%%%%
\section{Main theorem}
\label{sec:main-theorem}

In this section we present the main theorem.
%%%%%%%%%%%%%%%%%%%%%%%%%%%%%%%%%%%%%%%%%%%%%%%%%%%%%%%%%%%%%%%%%%%%
\vspace{1em} \\
\noindent {\bf Theorem 1 }{\it 
  Let Hamiltonian be represented as follows:
%%%%
  \begin{equation}
    \label{Hamiltonian}
        H(q_1,q_2,p_1,p_2) = H_0(q_1,q_2,p_1,p_2) +  \epsilon V(q_1,q_2),
  \end{equation} 
%%%%
where
%%%%
  \begin{equation}
        H_0(q_1,q_2,p_1,p_2) = \F{1}{2}(p_1^2 + p_2 ^2 + q_1^2 + q_2^2),
  \end{equation}
%%%%
  $\epsilon$ is a small parameter and 
  the potential $V(q_1,q_2)$ is an analytic function
  which has only even order terms of $q_1$ and $q_2$,
  for instance, $q_1 q_2^3$.
  We perturbatively solve the equation of motion
  by setting the $0$-th order solution as
%%%%
  \begin{equation}
    \label{0th-sol}
        q_{j0}(t) = B_j \cos t + C_j \sin t. \quad (j=1,2)
  \end{equation}
%%%%
  Then the following two conditions (A) and (B) are equivalent. \\
  (A) The renormalization group equation (RGE) of the system
  (\ref{Hamiltonian}) is a Hamiltonian system
  up to the second order of $\epsilon$. \\
  (B) The original system (\ref{Hamiltonian}) is separable
  by Cartesian coordinates, 
  that is, rotation of the coordinates $q_1$ and $q_2$. \\
  Moreover, when (A) (or (B)) is satisfied,
  the RGE is integrable and $(B_1^2+C_1^2+B_2^2+C_2^2)/2$ 
  is an integral which corresponds to $H_0$.
}\\
%%%%%%%%%%%%%%%%%%%%%%%%%%%%%%%%%%%%%%%%%%%%%%%%%%%%%%%%%%%%%%%%%%%%

\noindent
Remark 1: RGE for the system (\ref{Hamiltonian}) 
        is always a Hamiltonian system up to $O(\epsilon^1)$.
        Hamiltonian of the RGE is the same 
        as one obtained by canonical perturbation theory
        \cite{hori-66,deprit-69} 
        and the Hamiltonian is time-average of $V(q_{10},q_{20})$
        as shown in subsection \ref{sec:Cond-H}. \\

\noindent
Remark 2: RGE is an equation for variables which are constants of integration
        introduced in the $0$-th order solution,
        and hence the form of RGE changes  when we use
        different form of the $0$-th order solution, for instance,
        \begin{equation}
                \label{different-0th}
                q_{j0}(t) = \sqrt{2 A_j} \cos(t+\theta_j). 
        \end{equation}
        However, the theorem holds for  new variables 
        if they are canonical conjugate pair, 
        since constructing RGE and changing constants of integration
        are commutable operations.  \\

\noindent
Remark 3: We excluded odd order terms of $q_1$ and $q_2$ 
        from the perturbation $V(q_1,q_2)$.
        We must construct RGE up to $O(\epsilon^4)$
        if $V(q_1,q_2)$ has odd order terms
        (c.f. \cite{yamaguchi-98b}) and need longer calculation. But the
 extension is straightforward.\\

We prove this theorem 1 in sections \ref{sec:AtoB}, \ref{sec:BtoA}
 and \ref{sec:integral}. Sections \ref{sec:AtoB} and \ref{sec:BtoA}
are devoted to prove (A)$\Rightarrow$(B) and (B)$\Rightarrow$(A),
respectively. Section \ref{sec:integral} is for integrability and 
the form of an integral.

%%%%%%%%%%%%%%%%%%%%%%%%%%%%%%%%%%%%%%%%%%%%%%%%%%%%%%%%%%%%%%%
%% Section 4
%%%%%%%%%%%%%%%%%%%%%%%%%%%%%%%%%%%%%%%%%%%%%%%%%%%%%%%%%%%%%%%
\section{Proof of (A) $\Longrightarrow$ (B)}
\label{sec:AtoB}

We first derive RGE up to $O(\epsilon^2)$ 
for system (\ref{Hamiltonian}),
and we investigate conditions that RGE is a Hamiltonian system.
Then we compare the conditions with
separability conditions of system (\ref{Hamiltonian}).

%%%%%%%%%%%%%%%%%%%%%%%%%%%%%%%%%%%%%%%%%%%%%%%%%%%%%
\subsection{Renormalization Group Equation}
\label{sec:RGE-two}

Equation of motion in system (\ref{Hamiltonian}) is
%%%%
\begin{equation}
\eqalign{
        \ddot{q}_1 + q_1 &= - \epsilon V_{1}(q_1,q_2), \\
        \ddot{q}_2 + q_2 &= - \epsilon V_{2}(q_1,q_2),
}
\end{equation}
%%%%
where the subscripts of $V$ represent partial derivatives 
with the variables, namely,
$V_{1}=\partial V/\partial q_1$ and
$V_{12}=\partial^2 V/\partial q_1\partial q_2$.
We expand $q_1$ and $q_2$ as power series of $\epsilon$,
%%%%
\begin{equation}
\eqalign{
        q_1 &= q_{10} + \epsilon q_{11} + \epsilon^2 q_{12} + \cdots, \\
        q_2 &= q_{20} + \epsilon q_{21} + \epsilon^2 q_{22} + \cdots.
}
\end{equation}
%%%%
Equations of motion in each order of $\epsilon$ are
%%%%
\begin{eqnarray}
        \label{eq-motion-0}
        \ddot{q}_{j0} + q_{j0} &=& 0, \\
        \label{eq-motion-1}
        \ddot{q}_{j1} + q_{j1} &=& - V_{j}(\QZ), \\
        \label{eq-motion-2}
        \ddot{q}_{j2} + q_{j2} &=& - \sum_{k=1}^{2} V_{jk}(\QZ)q_{k1},
\end{eqnarray}
%%%%
where $\QZ=(q_{10},q_{20})$ and $j=1,2$.
Following an assumption described in theorem 1,
we set the $0$-th order solution as
%%%%
\begin{equation}
        %\label{0th-sol}
        q_{j0} = B_j \cos t + C_j \sin t. \quad (j=1,2)
\end{equation}
%%%%
Since  $V_{j}(\QZ)$ is a periodic function of 
time with period $2\pi$,
we expand it by Fourier series:
%%%%
\begin{equation}
        \label{Fourier-1}
        - V_{j}(\QZ) = \U{j}{0} + \sum_{n=1}^{\infty} \left(
                \U{j}{n}\cos nt + \V{j}{n}\sin nt \right) .             
\end{equation}
%%%%
Then solution to $O(\epsilon^1)$ is
%%%%
\begin{equation}
        \label{1st-sol}
 \hspace*{-3em}
        q_{j1} = \U{j}{0} 
        + \F{(t-t_0)}{2} (\U{j}{1} \sin t - \V{j}{1} \cos t)
        + \sum_{n=2}^{\infty} \F{1}{1-n^2}
                (\U{j}{n} \cos nt + \V{j}{n} \sin nt),
\end{equation}
%%%%
where $t_0$ is the initial time.

Let us proceed to solution to $O(\epsilon^2)$.
We also expand $V_{jk}(\QZ)$ by Fourier series,
%%%%
\begin{equation}
        \label{Fourier-2}
        - V_{jk}(\QZ) = \U{jk}{0} + \sum_{n=1}^{\infty}
                (\U{jk}{n}\cos nt + \V{jk}{n}\sin nt).
\end{equation}
%%%%
Terms in the right-hand-side of equation (\ref{eq-motion-2}) 
which make secular terms in $q_{j2}$ are
%%%%
\begin{equation}
        - \sum_{k=1}^{2} V_{jk}(\QZ)q_{k1} = 
        G_{j} t \cos t + H_{j} t \sin t
        + K_{j} \cos t + J_{j} \sin t, 
\end{equation}
%%%%
where, by using relations between Fourier components shown 
in \ref{app:relations},
%%%%
\begin{eqnarray}
        \label{G_j}
 \hspace*{-5em}
        G_j &=& 2 \sum_{k=1}^{2} \left( \fracp{\uz}{B_k} 
                \frac{\partial^2 \uz}{\partial B_j \partial C_k}
                - \fracp{\uz}{C_k} 
                \frac{\partial^2 \uz}{\partial B_j \partial B_k} \right), \\ 
        \label{H_j}
 \hspace*{-5em}
        H_j &=& 2 \sum_{k=1}^{2} \left( \fracp{\uz}{B_k} 
                \frac{\partial^2 \uz}{\partial C_j \partial C_k}
                - \fracp{\uz}{C_k} 
                \frac{\partial^2 \uz}{\partial C_j \partial B_k} \right), \\
        \label{K_j}
 \hspace*{-5em}
        K_j &=& \F{\partial}{\partial B_j} \left\{
                (\U{1}{0})^2 + (\U{2}{0})^2
                + \sum_{n=2}^{\infty} \F{1}{1-n^2}
                \left[ (\U{1}{n})^2 + (\V{1}{n})^2 
                + (\U{2}{n})^2 + (\V{1}{n})^2 \right] \right\}, \\
        \label{J_j}
 \hspace*{-5em}
        J_j &=& \F{\partial}{\partial C_j} \left\{
                (\U{1}{0})^2 + (\U{2}{0})^2
                + \sum_{n=2}^{\infty} \F{1}{1-n^2}
                \left[ (\U{1}{n})^2 + (\V{1}{n})^2 
                + (\U{2}{n})^2 + (\V{1}{n})^2 \right] \right\}.
\end{eqnarray}
%%%%
Solution to $O(\epsilon^2)$ is written with these symbols as
%%%%
\begin{eqnarray}
        \label{2nd-sol}
        q_{j2} &=& (t-t_0) \left[ 
                \left(\F{G_j}{4}-\F{J_j}{2}\right)\cos t
                + \left(\F{H_j}{4}+\F{K_j}{2}\right)\sin t \right] \NO
               &&  + (t-t_0)^2 \left[ 
                -\F{H_j}{4}\cos t + \F{G_j}{4}\sin t \right] 
              + \mbox{(non-secular terms)}.
\end{eqnarray}
%%%%
From the secular terms of equations (\ref{1st-sol}) and (\ref{2nd-sol}),
we obtain RGE as
%%%%
\begin{equation}
  \label{RGE}
\eqalign{
        \dot{B_j} &= - \F{\epsilon}{2}\V{j}{1}
                        + \epsilon^2 \left(\F{G_j}{4}-\F{J_j}{2}\right)
                = \fracp{\Phi}{C_j} + \F{\epsilon^2}{4} G_j, \\
        \dot{C_j} &=   \F{\epsilon}{2}\U{j}{1}
                        + \epsilon^2 \left(\F{H_j}{4}+\F{K_j}{2}\right)
                = -\fracp{\Phi}{B_j} + \F{\epsilon^2}{4} H_j, 
}
\end{equation}
%%%%
where the potential $\Phi$ is defined by
%%%%
\begin{equation}
 \hspace*{-5em}
        \Phi = - \epsilon \U{}{0} - \F{\epsilon^2}{2}
                \left\{  (\U{1}{0})^2 + (\U{2}{0})^2
                + \sum_{n=2}^{\infty} \F{1}{1-n^2}
                [ (\U{1}{n})^2 + (\V{1}{n})^2 
                + (\U{2}{n})^2 + (\V{2}{n})^2 ] \right\},
\end{equation}
%%%%
and $\uz$ is the constant component of Fourier series of $-V(\QZ)$
%%%%
\begin{equation}
        \uz = - \F{1}{2\pi} \int_0^{2\pi} V(q_{10},q_{20}) \ \DT.
\end{equation}
%%%%%%%%%%%%%%%%%%%%%%%%%%%%%%%%%%%%%%%%%%%%%%%%%%%%%%%%%%%%%%%%%

\subsection{Conditions that RGE is a Hamiltonian system}
\label{sec:Cond-H}

We show the conditions that the RGE (\ref{RGE}) becomes
a Hamiltonian system.
The conditions are
%%%%
\begin{equation}
        \label{conditions}
        \Delta_1=\Delta_2=\Delta_3=\Delta_4=\Delta_5=\Delta_6=0,
\end{equation}
%%%%
where  $\Delta_l\ (l=1,2,\cdots,6)$ are defined by
%%%%
\begin{equation}
\eqalign{
        \Delta_1 \equiv \fracp{\dot{B}_1}{B_1} + \fracp{\dot{C}_1}{C_1}, \quad
        \Delta_2 \equiv \fracp{\dot{B}_2}{B_2} + \fracp{\dot{C}_2}{C_2}, \quad
        \Delta_3 \equiv \fracp{\dot{C}_1}{B_2} - \fracp{\dot{C}_2}{B_1}, \\
        \Delta_4 \equiv \fracp{\dot{B}_1}{C_2} - \fracp{\dot{B}_2}{C_1}, \quad
        \Delta_5 \equiv \fracp{\dot{B}_1}{B_2} + \fracp{\dot{C}_2}{C_1}, \quad
        \Delta_6 \equiv \fracp{\dot{C}_1}{C_2} + \fracp{\dot{B}_2}{B_1}.
}
\end{equation}
%%%%
The term written by $\Phi$ in equation (\ref{RGE}) always satisfies
conditions (\ref{conditions}).
Accordingly, whether RGE is a Hamiltonian system is determined
by $G_j$ and $H_j$, 
which originates in secular terms of the $1$-st order solution. 
From equations (\ref{G_j}) and (\ref{H_j}),
the concrete forms of $\Delta_l$'s become
%%%%
\begin{equation}
\eqalign{
        \label{Deltas}
\fl
 \Delta_1 = \F{\epsilon^2}{2} \sum_{k=1}^{2} 
                   \left( \fracp{\uz}{B_k} \fracp{}{C_k}
                     - \fracp{\uz}{C_k} \fracp{}{B_k} \right)
                \left( \F{\partial^2}{\partial B_1^2}
                        + \F{\partial^2}{\partial C_1^2} \right) \uz~, \\
\fl
  \Delta_2 =  \F{\epsilon^2}{2} \sum_{k=1}^{2} 
                 \left( \fracp{\uz}{B_k} \fracp{}{C_k}
                     - \fracp{\uz}{C_k} \fracp{}{B_k} \right)
                \left( \F{\partial^2}{\partial B_2^2}
                        + \F{\partial^2}{\partial C_2^2} \right) \uz~, \\ 
\fl
  \Delta_3 = \Delta_4  \ =\  \F{\epsilon^2}{2} \Bigg[
                \frac{\partial^2 \uz}{\partial B_1 \partial B_2}
                \left( \frac{\partial^2}{\partial C_1^2}
                -\frac{\partial^2}{\partial C_2^2} \right) \uz 
                 -\frac{\partial^2 \uz}{\partial C_1 \partial C_2}
                \left( \frac{\partial^2}{\partial B_1^2}
                -\frac{\partial^2}{\partial B_2^2} \right) \uz \Bigg]~, \\
\fl
   \Delta'_5 =  \F{\epsilon^2}{2} 
                        \left( \fracp{\uz}{B_1} \fracp{}{C_1}
                        -\fracp{\uz}{C_1} \fracp{}{B_1}
                        +\fracp{\uz}{B_2} \fracp{}{C_2}
                        -\fracp{\uz}{C_2} \fracp{}{B_2}  \right) \\
                    \hspace*{8em}
                \times \left( \frac{\partial^2}{\partial B_1 \partial B_2}
                +\frac{\partial^2}{\partial C_1 \partial C_2} \right) \uz~, \\
\fl
   \Delta'_6=  \F{\epsilon^2}{2} \left[
                        \left( \frac{\partial^2}{\partial B_1 \partial B_2}
                        +\frac{\partial^2}{\partial C_1 \partial C_2} \right)
                        \uz \cdot
                      \left( \frac{\partial^2}{\partial B_1 \partial C_1}
                        -\frac{\partial^2}{\partial B_2 \partial C_2} \right) 
                        \uz \right. \\
                   \hspace*{5em} 
                \left.
                - \frac{\partial^2 \uz}{\partial B_1 \partial C_2}
                \left( \frac{\partial^2}{\partial B_1^2}
                  +\frac{\partial^2}{\partial C_1^2}
                  -\frac{\partial^2}{\partial B_2^2}
                  -\frac{\partial^2}{\partial C_2^2} \right) \uz 
                \right] ,
}
\end{equation}
%%%%
where
%%%%
\begin{equation}
\eqalign{
        \Delta_5 &= \Delta'_5 + \Delta'_6 ~, \\
        \Delta_6 &= \Delta'_5 - \Delta'_6 ~.
}
\end{equation}
%%%%

Note that RGE is always a Hamiltonian system up to $O(\epsilon^1)$
since terms of $O(\epsilon^1)$ do not appear in $\Delta_l$'s.
 In this order, a Hamiltonian of the RGE becomes time-average of
$V(\QZ)$:
%%%%
\begin{equation}
        \HRG_1 = -\uz = \F{1}{2\pi}\int_{0}^{2\pi} \DT \ V(\QZ),
\end{equation}
%%%%
and $\HRG_1$ is the same as the one obtained by canonical
perturbation theory \cite{hori-66,deprit-69}.

%%%%%%%%%%%%%%%%%%%%%%%%%%%%%%%%%%%%%%%%%%%%%%%%%%%%%%%%%%%%
\subsection{Condition that the original system is separable}
\label{sec:separable}

In this subsection we clarify the condition
that the original system (\ref{Hamiltonian}) is separable
by rotation of the coordinates $(q_1,q_2)$.
The integrable part $H_0$ is always separable by any rotation,
and hence we only have to consider the perturbation $V$.
In the next subsection, we compare the condition written by $\uz$
with equation (\ref{Deltas}). \\
%%%%%%%%%%%%%%%%%%%%%%%%%%%%%%%%%%%%%%%%%%%%%%%%%%%%%%%%%%%%%%%%%%%%

\noindent 
{\bf Theorem 2 }{\it 
        The condition that the original system (\ref{Hamiltonian})
        becomes separable by rotation of the coordinates 
        $(q_1,q_2)$ is
        \begin{equation}
                \label{separability}
                c_1 (V_{11}-V_{22}) + 2 c_2 V_{12} = 0. 
                \quad (c_1^2+c_2^2 =1)
        \end{equation}
        In particular, when the potential $V(q_1,q_2)$ has only
        even order terms of $q_1$ and $q_2$,
        the condition (\ref{separability}) is equivalent 
        to each of the following three conditions:
        \begin{eqnarray}
        \hspace*{-4em} \mbox{(I)} &&
        \left[ c_1 \left(       
                  \frac{\partial^2}{\partial B_1^2} 
                + \frac{\partial^2}{\partial C_1^2} 
                - \frac{\partial^2}{\partial B_2^2} 
                - \frac{\partial^2}{\partial C_2^2} \right)
         + 2 c_2 \left( \frac{\partial^2}{\partial B_1 \partial B_2}
                +  \frac{\partial^2}{\partial C_1 \partial C_2} \right) 
                \right] \uz = 0, \\
        \hspace*{-4em}\mbox{(II)} && 
        \left[ c_1 \left(       
                  \frac{\partial^2}{\partial B_1^2} 
                - \frac{\partial^2}{\partial C_1^2} 
                - \frac{\partial^2}{\partial B_2^2} 
                + \frac{\partial^2}{\partial C_2^2} \right)
         + 2 c_2 \left( \frac{\partial^2}{\partial B_1 \partial B_2}
                -  \frac{\partial^2}{\partial C_1 \partial C_2} \right) 
                \right] \uz = 0, \\
        \hspace*{-4em}\mbox{(III)} && 
        \left[ c_1 \left(
                \frac{\partial^2}{\partial B_1 \partial C_1}
                - \frac{\partial^2}{\partial B_2 \partial C_2} \right)
        + c_2 \left(
                \frac{\partial^2}{\partial B_1 \partial C_2}
                + \frac{\partial^2}{\partial C_1 \partial B_2} \right) 
                \right] \uz = 0. 
        \end{eqnarray}
}\\

%%%%%%%%%%%%%%%%%%%%%%%%%%%%%%%%%%%%%%%%%%%%%%%%%%%%%%%%%%%%%%%%%%%%
\noindent 
{\bf Corollary 3 }{\it 
From (I) and (II), the two conditions
\begin{eqnarray}
        \mbox{(I')} &&
        \left[ c_1 \left(       
                  \frac{\partial^2}{\partial B_1^2} 
                - \frac{\partial^2}{\partial B_2^2} \right)
         + 2 c_2 \left( 
                \frac{\partial^2}{\partial B_1 \partial B_2} \right)
                \right] \uz = 0, \\
        \mbox{(II')} &&
        \left[ c_1 \left(       
                  \frac{\partial^2}{\partial C_1^2} 
                - \frac{\partial^2}{\partial C_2^2} \right)
         + 2 c_2 \left( 
                \frac{\partial^2}{\partial C_1 \partial C_2} \right)
                \right] \uz = 0,
\end{eqnarray}
are also equivalent to each of (I), (II) and (III).
}\\

%%%%%%%%%%%%%%%%%%%%%%%%%%%%%%%%%%%%%%%%%%%%%%%%%%%%%%%%%%%%%%%%%%%%

Equation (\ref{separability}) is a special case
of Darboux equation \cite{hietarinta-87},
which is used in the previous paper \cite{yamaguchi-98b}.
To prove theorem 2, we introduce the following lemma.\\

\noindent 
{\bf Lemma 4 }{\it 
        Let us assume that $V(q_1,q_2)$ has only even order terms of 
        $q_1$, $q_2$ and $V(q_1,q_2)$ is not constant. 
        We define the function $F(q_1,q_2)$ as
        \begin{equation}
                F(q_1,q_2) \equiv c_1 [ V_{11}(q_1,q_2) - V_{22}(q_1,q_2) ]
                                + 2 c_2 V_{12}(q_1,q_2),
        \end{equation}
        then
        \begin{eqnarray}
                && \mbox{(a)} \quad F(q_1,q_2)=0 \quad \mbox{for }
                {}^\forall{q}_1, {}^\forall{q}_2 \\
                &\Leftrightarrow& \mbox{(b)} \quad
                \int_0^{2\pi} \DT F(q_{10},q_{20}) = 0
                \quad \mbox{for } 
                {}^\forall{B}_1, {}^\forall{C}_1, {}^\forall{B}_2, {}^\forall{C}_2
\\
                &\Leftrightarrow& \mbox{(c)} \quad
                \int_0^{2\pi} \DT F(q_{10},q_{20}) \cos 2t = 0
                \quad \mbox{for } 
                {}^\forall{B}_1, {}^\forall{C}_1, {}^\forall{B}_2, {}^\forall{C}_2
\\
                &\Leftrightarrow& \mbox{(d)} \quad
                \int_0^{2\pi} \DT F(q_{10},q_{20}) \sin 2t = 0
                \quad \mbox{for } 
                {}^\forall{B}_1, {}^\forall{C}_1, {}^\forall{B}_2, {}^\forall{C}_2
        \end{eqnarray}
}\\
%%%%%%%%%%%%%%%%%%%%%%%%%%%%%%%%%%%%%%%%%%%%%%%%%%%%%%%%%%%%%%%%%%%%

\noindent 
{\bf Proof of Lemma 4:} \\
The direction (a) $\Rightarrow$ (b),(c),(d) is obvious,
and hence we prove the reverse direction,
that is, (b),(c),(d) $\Rightarrow$ (a).
We may set $C_1=C_2=0$ 
since (b),(c) and (d) are satisfied 
for any values of $B_j$ and $C_j$.
Then $F(q_{10},q_{20})$ is expanded as
%%%%
\begin{equation}
        F(q_{10},q_{20}) = \sum_{m=0}^{\infty} \sum_{n=0}^{\infty}
                \frac{F^{(m,n)}(0,0)}{m!n!} B_{1}^{m} B_{2}^{n}
                \cos^{m+n} t,
\end{equation}
%%%%
where
%%%%
\begin{equation}
        F^{(m,n)}(0,0)
        =\F{\partial^{m+n} F}{\partial q_{1}^{m} \partial q_{2}^{n}}
        (0,0).
\end{equation}
%%%%
As a result, in case of (b),
%%%%
\begin{eqnarray}
 \hspace*{-3em}
        0 &=& \int_{0}^{2\pi} F(q_{10},q_{20}) \DT \NO
 \hspace*{-3em}
        &=& \sum_{m=0}^{\infty} \sum_{n=0}^{\infty}
                \frac{F^{(m,n)}(0,0)}{m!n!} B_{1}^{m} B_{2}^{n}\times
                \left\{
        \begin{array}{ll}
                2\pi{2^{-m-n}} {}_{(m+n)}C_{(m+n)/2} &
                ~(m+n: \mbox{even}) \\
                0 & ~ (m+n: \mbox{odd}) 
        \end{array}
                \right.
\end{eqnarray}
%%%%
and $F^{(m,n)}(0,0)=0$ for even $m+n$.
Consequently, $F(q_{1},q_{2})=0$ because $F(q_{1},q_{2})$ has only 
even order terms of $q_{1}$ and $q_{2}$.
We can prove $(c),(d) \Rightarrow (a)$ through similar ways.
$\Box$\\

Now let us prove theorem 2.\\

\noindent 
{\bf Proof of Theorem 2:}
Let us prove the first half of theorem 2.
We rotate coordinates from $(q_1,q_2)$ to $(\tilde{q}_1,\tilde{q}_2)$ as
%%%%
\begin{equation}
 \label{rotation}
\eqalign{
        q_1 &= \tilde{q}_1 \cos \theta + \tilde{q}_2 \sin \theta, \\ 
        q_2 &= - \tilde{q}_1 \sin \theta + \tilde{q}_2 \cos \theta.
}
\end{equation}
%%%%
The separability condition is that there exists 
a real value of $\theta$ such that
%%%%
\begin{equation}
        \frac{\partial^2 V}{\partial \tilde{q}_1 \partial \tilde{q}_2} = 0,
        \quad \mbox{for } {}^\forall{\tilde{q}}_1,{}^\forall{\tilde{q}}_2.
\end{equation}
%%%%
This condition is equivalent to
%%%%
\begin{equation}
        \label{separability-2}
        (V_{11}-V_{22}) \sin 2\theta + 2 V_{12} \cos 2\theta = 0,
        \quad \mbox{for } {}^\forall{q}_1,{}^\forall{q}_2,
\end{equation}
%%%%
and consequently, we have proved the first half of theorem 2.

From relations between Fourier components
shown in \ref{app:relations},
the conditions (I),(II) and (III) are equivalent to
%%%%
\begin{eqnarray}
        \mbox{(I)}: 
                & & \int_{0}^{2\pi} \DT F(q_{10},q_{20}) = 0, \\
        \mbox{(II)}: 
                & & \int_{0}^{2\pi} \DT F(q_{10},q_{20}) \cos 2t = 0, \\
        \mbox{(III)}: 
                & & \int_{0}^{2\pi} \DT F(q_{10},q_{20}) \cos t\sin t = 
                \F{1}{2} \int_{0}^{2\pi} \DT F(q_{10},q_{20}) \sin 2 t = 0,
\end{eqnarray}
%%%%
respectively.
Then, from lemma 4, we conclude that theorem 2 has been proved.
$\Box$

%%%%%%%%%%%%%%%%%%%%%%%%%%%%%%%%%%%%%%%%%%%%%%%%%%%%%%%%%%%%%%%%%%%%%
\subsection{Final step of proof of (A)$\Rightarrow$(B)}
\label{sec:finalAtoB}

We prove (A)$\Rightarrow$(B) of theorem 1
with the aid of theorem 2 and corollary 3.
The conditions (I') and (II') are rewritten in matrix form
%%%%
\begin{equation}
        \left( 
        \begin{array}{cc}
%                (\Par{B_1}^2-\Par{B_2}^2) \uz &
%                2\Par{B_1}\Par{B_2} \uz \\
%                (\Par{C_1}^2-\Par{C_2}^2) \uz &
%                2\Par{C_1}\Par{C_2} \uz
          (\frac{\partial^2}{\partial B_1^2}
          - \frac{\partial^2}{\partial B_2^2}) \uz &
          2 \frac{\partial^2}{\partial B_1 \partial B_2} \uz \\
          (\frac{\partial^2}{\partial C_1^2}
          - \frac{\partial^2}{\partial C_2^2}) \uz &
          2 \frac{\partial^2}{\partial C_1 \partial C_2} \uz
        \end{array}
        \right)
        \left( 
        \begin{array}{c}
                c_1 \\
                c_2
        \end{array}
        \right)
        =
        \left( 
        \begin{array}{c}
                0 \\
                0
        \end{array}
        \right).
\end{equation}
%%%%
The constants $c_1$ and $c_2$ are not zero simultaneously,
and the conditions (I') and (II') must be degenerate accordingly.
The degenerate condition is
%%%%
\begin{eqnarray}
 \hspace*{-3em}
        && \mbox{det} \left(
        \begin{array}{cc}
%                (\Par{B_1}^2-\Par{B_2}^2) \uz &
%                2\Par{B_1}\Par{B_2} \uz \\
%                (\Par{C_1}^2-\Par{C_2}^2) \uz &
%                2\Par{C_1}\Par{C_2} \uz
          (\frac{\partial^2}{\partial B_1^2}
          - \frac{\partial^2}{\partial B_2^2}) \uz &
          2 \frac{\partial^2}{\partial B_1 \partial B_2} \uz \\
          (\frac{\partial^2}{\partial C_1^2}
          - \frac{\partial^2}{\partial C_2^2}) \uz &
          2 \frac{\partial^2}{\partial C_1 \partial C_2} \uz
        \end{array}
        \right)=0 \NO
 \hspace*{-3em}
        && \Longleftrightarrow
        \frac{\partial^2 \uz}{\partial B_1 \partial B_2}
                \left( \frac{\partial^2}{\partial C_1^2}
        - \frac{\partial^2}{\partial C_2^2} \right) \uz 
                - \frac{\partial^2 \uz}{\partial C_1 \partial C_2}
        \left( \frac{\partial^2}{\partial B_1^2}
                -\frac{\partial^2}{\partial B_2^2} \right) \uz
        = 0,
\end{eqnarray}
%%%%
and this separable condition is equivalent to $\Delta_3=\Delta_4=0$.
We also find that the degenerate condition between (I) and (III) 
is equivalent to $\Delta_6=0$.
Consequently,
%%%%
\begin{eqnarray*}
        && \hspace*{-3em}
        \mbox{RGE is a Hamiltonian system up to } O(\epsilon^2) \\
        &\Longleftrightarrow& \Delta_1=\Delta_2=\Delta_3=
                \Delta_4=\Delta_5=\Delta_6=0 \\
%        &\Longrightarrow& \Delta_3=\Delta_4=\Delta_6=0 \\
        &\Longrightarrow& \mbox{Potential function } V(q_1,q_2) 
                \mbox{ is separable by Cartesian coordinates.} 
\end{eqnarray*}
%%%%%%%%%%%%%%%%%%%%%%%%%%%%%%%%%%%%%%%%%%%%%%%%%%%%%%%%%%%%%%%
%% Section 5
%%%%%%%%%%%%%%%%%%%%%%%%%%%%%%%%%%%%%%%%%%%%%%%%%%%%%%%%%%%%%%%
\section{Proof of (B) $\Longrightarrow$ (A)}
\label{sec:BtoA}

We assume that the Hamiltonian system (\ref{Hamiltonian}) 
is separable by Cartesian coordinates.
RGE for the Hamiltonian system is also separable
since constructing RGE and rotating coordinates 
are commutable operations.
As a result, what we have to prove is the following theorem.\\

%%%%%%%%%%%%%%%%%%%%%%%%%%%%%%%%%%%%%%%%%%%%%%%%%%%%%%%%%%%%%%%
\noindent 
{\bf Theorem 5 }{\it 
 RGE  up to $O(\epsilon^2)$
for the system with one degree of freedom
%%%%
\begin{equation}
        \label{Hamiltonian_1}
        H(q_1,p_1) = \F{1}{2} (p_1^2 + q_1^2) + \epsilon V (q_1)
\end{equation}
%%%%
is a Hamiltonian system. $V$ is an analytic function of $q_1$.
}\\

\noindent 
{\bf Proof of Theorem 5: } 
RGE for system (\ref{Hamiltonian_1}) is obtained
from equation (\ref{RGE}) by omitting terms with subscript $2$:
%%%%
\begin{equation}
  \label{RGE-one-degree}
\eqalign{
        \dot{B_1} &= \fracp{\Phi}{C_1}
                        + \epsilon^2 \F{G_1}{4},\\
        \dot{C_1} &= - \fracp{\Phi}{B_1}
                        + \epsilon^2 \F{H_1}{4},
}
\end{equation}
%%%%
where
%%%%
\begin{eqnarray}
        \Phi &=& - \epsilon \U{}{0} - \F{\epsilon^2}{2}
                \left\{  (\U{1}{0})^2 
                + \sum_{n=2}^{\infty} \F{1}{1-n^2}
                [ (\U{1}{n})^2 + (\V{1}{n})^2 ] \right\}, \\ 
        \label{G_1}
        G_1 &=& 2 \left(
                \fracp{\uz}{B_1} 
                \frac{\partial^2 \uz}{\partial B_1 \partial C_1}
                -\fracp{\uz}{C_1} 
                \frac{\partial^2 \uz}{\partial B_1^2} \right), \\
        \label{H_1}
        H_1 &=& 2  \left(
                \fracp{\uz}{B_1} 
                \frac{\partial^2 \uz}{\partial C_1^2}
                -\fracp{\uz}{C_1} 
                \frac{\partial^2 \uz}{\partial B_1 \partial C_1}
                 \right).
\end{eqnarray}
%%%%
The condition that the RGE (\ref{RGE-one-degree}) becomes 
a Hamiltonian system is
%%%%
\begin{equation}
        \Delta_1 \equiv \fracp{\dot{B}_1}{B_1} + \fracp{\dot{C_1}}{C_1} = 0,
\end{equation}
%%%%
and the  form of $\Delta_1$ is
%%%%
\begin{eqnarray}
        \F{4}{\epsilon^2} \Delta_1 
        &=& \fracp{G_1}{B_1} + \fracp{H_1}{C_1} \NO
        &=& 2 \left[
        \fracp{\uz}{B_1} \frac{\partial}{\partial C_1}
        \left( \frac{\partial^2}{\partial B_1^2}
         + \frac{\partial^2}{\partial C_1^2} \right) \uz
        - \fracp{\uz}{C_1} \frac{\partial}{\partial B_1}
        \left( \frac{\partial^2}{\partial B_1^2}
         + \frac{\partial^2}{\partial C_1^2} \right) \uz 
        \right] \NO
        &=& \fracp{\uz}{B_1} \fracp{\U{11}{0}}{C_1}
        - \fracp{\uz}{C_1} \fracp{\U{11}{0}}{B_1} .
\end{eqnarray}
%%%%

We prove that the $\Delta_1$ is always zero
by showing both $\uz$ and $\U{11}{0}$ are functions of 
$B_1^2+C_1^2$.
From the definition of $\uz$,
%%%%
\begin{equation}
\label{uzero}
\eqalign{
   \uz &= -\frac{1}{2\pi}\int_0^{2\pi}dt V(B_1\cos t+C_1\sin t) \\
        &=-\frac{1}{2\pi}\int_0^{2\pi}dt
V(\sqrt{B_1^2+C_1^2}\cos(t-\delta))  \\
&=-\frac{1}{2\pi}\sum_{n=0}^{\infty}\frac{V^{(n)}(0)}{n!}
   (B_1^2+C_1^2)^{n/2}\int_0^{2\pi}dt\cos^n(t-\delta) \\
&=-\sum_{n=0}^{\infty}\frac{V^{(2n)}(0)}{(2n)!}
   (B_1^2+C_1^2)^{n}\times 2^{-2n}{}_{2n}C_{n},
}
\end{equation}
%%%%
where $\tan\delta=C_1/B_1$ and $V^{(k)}= 
\partial^k V/\partial q^k$. 
Consequently,
$\uz$ is a function of $B_1^2+C_1^2$.
We can also prove that $\U{11}{0}$ is a function of $B_1^2+C_1^2$
through the same way.
$\Box$

%%%%%%%%%%%%%%%%%%%%%%%%%%%%%%%%%%%%%%%%%%%%%%%%%%%%%%%%%%%%%%%
%% Section 6
%%%%%%%%%%%%%%%%%%%%%%%%%%%%%%%%%%%%%%%%%%%%%%%%%%%%%%%%%%%%%%%
\section{Integral of renormalization group equation}
\label{sec:integral}

From the consideration  of the system with one degrees of 
freedom, RGE with two degrees of freedom is always integrable when it is
a Hamiltonian system since it is separable.
Moreover, $H_0$ of system (\ref{Hamiltonian}) 
is  an integral of RGE.\\

%%%%%%%%%%%%%%%%%%%%%%%%%%%%%%%%%%%%%%%%%%%%%%%%%%%%%%%%%%%%%%%%%%%%
\noindent 
{\bf Theorem 6 }{\it For the RGE system with one 
degrees of freedom (\ref{RGE-one-degree}),
        the quantity $B_{1}^{2}+C_{1}^{2}$ is an integral of 
        the equation. For the RGE system with two degrees of freedom
(\ref{RGE}),
        if RGE is a Hamiltonian system then 
$B_{1}^{2}+C_{1}^{2}+B_{2}^{2}+C_{2}^{2}$
        is an integral of the equation.
}\\
%%%%%%%%%%%%%%%%%%%%%%%%%%%%%%%%%%%%%%%%%%%%%%%%%%%%%%%%%%%%%%%%%%%%

\noindent 
{\bf Proof of Theorem 6: } 
First we consider systems with one degree of freedom.
We prove that the time derivative of $(B_{1}^{2}+C_{1}^{2})/2$ 
 becomes zero:
%%%%
\begin{eqnarray}
        \label{derivative}
        \F{1}{2}\F{\D}{\D t}(B_{1}^{2}+C_{1}^{2}) 
        &=& B_{1} \dot{B_{1}} + C_{1} \dot{C_{1}} \NO
        &=& B_{1} \fracp{\Phi}{C_{1}} - C_{1} \fracp{\Phi}{C_{1}}
                + \F{\epsilon^2}{4} (B_{1}G+C_{1}H) = 0.
\end{eqnarray} 
%%%%
We prove equation (\ref{derivative}) 
by showing that $\Phi$ is a function of $B_{1}^2+C_{1}^2$
and that $B_{1}G+C_{1}H=0$.

We first consider the terms written by $\Phi$. The
 concrete form of $\Phi$ is
%%%%
\begin{equation}
        \Phi = -\epsilon \U{}{0}
        - \F{\epsilon^2}{2} \left\{
                \U{1}{0} + \sum_{n=2}^{\infty} \F{1}{1-n^2}
                [ (\U{1}{n})^2 + (\V{1}{n})^2 ]
        \right\}.
\end{equation}
%%%%
The first and the second terms $\U{}{0}$, $\U{1}{0}$,
are constant components of $-V(q_0)$ and $-V_1(q_0)$,
respectively. 
They are therefore functions of $B_{1}^2+C_{1}^2$ accordingly
(c.f. equations (\ref{uzero})).
From the definition of $\U{1}{n}$,
%%%%
\begin{eqnarray}
        \U{1}{n} 
        &=& \frac{1}{\pi}\int_{0}^{2\pi} \DT
                V_{1}(B_{1}\cos t+C_{1}\sin t) \cos nt \NO
        &=&  \frac{1}{\pi}\int_{0}^{2\pi} \DT
                V_{1}(\sqrt{B_{1}^2+C_{1}^2}\cos (t-\delta)) \cos nt \NO
        &=&  \frac{1}{\pi}\int_{0}^{2\pi} \DT
                \sum_{k=0}^{\infty} \F{V_{1}^{(k)}(0)}{k!} (B_{1}^2+C_{1}^2)^{k/2} 
                \cos^k (t-\delta) \cos nt \NO
%       &=& \sum_{k=0}^{\infty} \F{V_{1}^{(k)}(0)}{k!} (B_{1}^2+C_{1}^2)^{k/2} 
%               \int_{0}^{2\pi} \DT \cos^k t \cos n(t+\delta) \NO
        &=& \frac{1}{\pi}\sum_{k=0}^{\infty} \F{V^{(k+1)}(0)}{k!}
(B_{1}^2+C_{1}^2)^{k/2} 
                ( C^{(k)}_{n} \cos n\delta 
                - S^{(k)}_{n} \sin n \delta ), 
\end{eqnarray}
%%%%
where 
%%%%
\begin{equation}
        C^{(k)}_{n} \equiv \int_{0}^{2\pi} \DT \cos^k t \cos nt, \qquad 
        S^{(k)}_{n} \equiv \int_{0}^{2\pi} \DT \cos^k t \sin nt.
\end{equation}
%%%%
We also have
%%%%
\begin{equation}
        \V{1}{n} = \frac{1}{\pi}\sum_{k=0}^{\infty} \F{V^{(k+1)}(0)}{k!} 
                (B_{1}^2+C_{1}^2)^{k/2} 
                ( C^{(k)}_{n} \sin n\delta 
                + S^{(k)}_{n} \cos n \delta ), 
\end{equation}
%%%%
and
%%%%
\begin{equation}
        \hspace*{-5.5em}
(\U{1}{n})^2 + (\V{1}{n})^2 
        = \frac{1}{\pi^2}\sum_{k=0}^{\infty} \sum_{l=0}^{\infty} 
                \F{V^{(k+1)}(0)V^{(l+1)}(0)}{k!\ l!} 
                (B_{1}^2+C_{1}^2)^{(k+l)/2} 
                ( C^{(k)}_{n} C^{(l)}_{n} + S^{(k)}_{n} S^{(l)}_{n} ).
\end{equation}
%%%%
Consequently, $\Phi$ is a function of $B_{1}^2+C_{1}^2$
since $C^{(k)}_{n} C^{(l)}_{n} + S^{(k)}_{n} S^{(l)}_{n}$ 
is independent of $B_{1}$ and $C_{1}$.

Next we prove that $B_{1}G+C_{1}H=0$.
From  (\ref{G_1}), (\ref{H_1}) and the fact
that $\uz$ is a function of $B_{1}^2+C_{1}^2$,
%%%%
\begin{equation}
        G =-2 C_{1} [(u^{(0)})']^2, ~~~H =  2 B_{1} [(u^{(0)})']^2,
\end{equation}
%%%%
where $(u^{(0)})'$ is derivative of $u^{(0)}(B_{1}^2+C_{1}^2)$,
and hence $B_{1}G+C_{1}H=0$.

Finally we consider systems with two degrees of freedom.
When RGE is a  Hamiltonian system,
both $B_1^2+C_1^2$ and $B_2^2+C_2^2$ are integrals of RGE
in separated coordinate.
Accordingly, $B_1^2+C_1^2+B_2^2+C_2^2$ is an integral
in any Cartesian coordinates
since it is invariant by rotation of the coordinates $(q_1,q_2)$
 which induces rotation of two pairs of variables: $(B_1,B_2)$ and
$(C_1,C_2)$.
$\Box$\\

Now we have completed proof of theorem 1.
$\Box$ 
%{\bf End of proof of Theorem 1}.
%%%%%%%%%%%%%%%%%%%%%%%%%%%%%%%%%%%%%%%%%%%%%%%%%%%%%%%%%%%%%%%
%% Section 7
%%%%%%%%%%%%%%%%%%%%%%%%%%%%%%%%%%%%%%%%%%%%%%%%%%%%%%%%%%%%%%%
\section{Summary and discussions}
\label{sec:summary}

We considered renormalization group equation (RGE)
for perturbed Hamiltonian system with two degrees of freedom
whose integrable part is two harmonic oscillators
with the same angular frequency,
and perturbation is a function of positions
having only even order terms.
RGE is not always Hamiltonian system 
and we presented the necessary and sufficient condition that 
 RGE becomes Hamiltonian system as the 
theorem 1. The theorem  states that RGE in Cartesian coordinates 
becomes a Hamiltonian system up to $O(\epsilon^2)$
if and only if the original system is separable by rotation
of the coordinates.
When RGE is a  Hamiltonian system,
it has an integral which corresponds to the integral part
of original system  
and RGE is integrable accordingly.
The theorem  also assert that RGE is a Hamiltonian system
when the original system has one degree of freedom.
  RGE is always Hamiltonian system
up to $O(\epsilon^1)$ 
and its Hamiltonian is the same as one obtained by 
canonical perturbation theory.

The theorem  excluded odd order terms from the perturbation
potential function.
We must construct RGE up to $O(\epsilon^4)$
to extend the theorem to any analytic perturbation functions
since odd order terms give information on the separability
at $O(\epsilon^4)$. But it is straightforward to extend our result 
to include odd order terms in potential function. 
We expect that the theorem holds for any perturbation
from the result of  homogeneous cubic perturbation case
\cite{yamaguchi-98b}, which satisfies the theorem.

RGE is not a Hamiltonian system 
when the original system  has two degrees of freedom 
and have no integral other than Hamiltonian of the system.
 On the other hand, RGE with one degree of freedom
 is  always Hamiltonian system.
We therefore expect that we can understand properties
of chaotic orbits by investigating 
how symplectic properties break
in the process of constructing RGE.
Solutions to non-Hamiltonian RGE will also give 
some aspects of chaos.

Future works of this topic is as follows.
We suppose that it is straightforward to extend the main theorem of 
this paper to systems with many degrees of freedom.
Another extension is generalization of integrable part 
of original systems.
We are particularly interested in 
 the case that integrable part is non-linear 
and angular frequencies change as values of actions.

%%%%%%%%%%%%%%%%%%%
% acknowledgement
%%%%%%%%%%%%%%%%%%%
\ack{
The authors would like to thank Hiroyasu Yamada and Atsushi Taruya for
teaching us techniques on singular perturbation and RGE.
Y.N. is supported in part by the Grand-In-Aid 
for Scientific Research of the Ministry of Education, Science, Sports and 
Culture of Japan(09740196).}

%%%
\newpage
%%%
\section*{References}
%%%%%%%%%%%%%%%%%%%%%%%%%%%%%%%%%%%%%%%%%%%%%%%%%%%%%%%%%%
% References
%%%%%%%%%%%%%%%%%%%%%%%%%%%%%%%%%%%%%%%%%%%%%%%%%%%%%%%%%%

%%%%%%%%%%%%%%%%%%%%%%%%%%%%%%%%%%%%%%%%%%%%%%%%%%%%%%%%%%%%%%%
\newpage
\appendix
%%%%%%%%%%%%%%%%%%%%%%%%%%%%%%%%%%%%%%%%%%%%%%%%%%%%%%%%%%%%%%%
%% Appendix A
%%%%%%%%%%%%%%%%%%%%%%%%%%%%%%%%%%%%%%%%%%%%%%%%%%%%%%%%%%%%%%%
\section{Some relations between Fourier components}
\label{app:relations}

We show some useful relations between Fourier components
of $V(\QZ)$, $-V_{j}(\QZ)$ and $-V_{jk}(\QZ)$.
 From the definition, $\U{1}{1}$ is represented by $\uz$ as
\begin{eqnarray}
        \U{1}{1} &=& \F{1}{\pi}\int_0^{2\pi} \DT 
                        V_{1}(q_{10},q_{20}) \cos t \NO
                &=& \fracp{}{B_1} \left[ \F{1}{\pi}\int_0^{2\pi} \DT
                        V(q_{10},q_{20}) \right] \NO
                &=& 2 \fracp{\U{}{0}}{B_1}.
\end{eqnarray}
Similar relations hold for other components:
%%%%
\begin{equation}
        \U{j}{1} = 2 \fracp{\U{}{0}}{B_j}, \quad
        \V{j}{1} = 2  \fracp{\U{}{0}}{C_j} ,         
\end{equation}
%%%%
and for $n\geq 2$,
%%%%
\begin{equation}
\eqalign{
        \U{j}{n-1} = \fracp{\U{}{n}}{B_j} + \fracp{\V{}{n}}{C_j}, \quad
        \U{j}{n+1} = \fracp{\U{}{n}}{B_j} - \fracp{\V{}{n}}{C_j}, \\
        \V{j}{n-1} = \fracp{\V{}{n}}{B_j} - \fracp{\U{}{n}}{C_j}, \quad
        \V{j}{n+1} = \fracp{\V{}{n}}{B_j} + \fracp{\U{}{n}}{C_j}, 
}
\end{equation}
%%%%
Some Fourier components of $V_{jk}(q_{10},q_{20})$ are
rewritten by $\uz$ as follows:
\begin{equation}
\eqalign{
        2 \U{jk}{0} &= \fracp{\U{k}{1}}{B_j} + \fracp{\V{k}{1}}{C_j}
                 = 2 \left( \F{\partial^2}{\partial B_j \partial B_k}
                     + \F{\partial^2}{\partial C_j \partial C_k} \right) 
                        \uz, \\
        \U{jk}{2} &=  \fracp{\U{k}{1}}{B_j} - \fracp{\V{k}{1}}{C_j}
                 = 2 \left( \F{\partial^2}{\partial B_j \partial B_k}
                     - \F{\partial^2}{\partial C_j \partial C_k} \right) 
                        \uz, \\
        \V{jk}{2} &=  \fracp{\V{k}{1}}{B_j} + \fracp{\U{k}{1}}{C_j}
                 = 2 \left( \F{\partial^2}{\partial B_j \partial C_k}
                     + \F{\partial^2}{\partial C_j \partial B_k} \right)
                        \uz.
}
\end{equation}
%%%%
From the definition of Fourier components,
the following equation is always satisfied 
for any Fourier components:
\begin{equation} 
        \F{\partial^2}{\partial B_1 \partial C_2} X
        = \F{\partial^2}{\partial C_1 \partial B_2} X, 
\end{equation} 
where $X$ is an arbitrary Fourier component.

%%%%%%%%%%%%%%%%%%%%%%%%%%%%%%%%%%%%%%%%%%%%%%%%%%%%%%%%%%%%%%%
%% Appendix B
%%%%%%%%%%%%%%%%%%%%%%%%%%%%%%%%%%%%%%%%%%%%%%%%%%%%%%%%%%%%%%%
\newpage
\section{Hamiltonian of RGE in quartic perturbation}
\label{app:HRG-quartic}

We show the explicit form of the Hamiltonian of RGE in quartic
perturbation when RGE is Hamiltonian system. First we describe how we
obtain the Hamiltonian from the RGE. Second the concrete form of the
Hamiltonian is shown.

\subsection{Method of constructing Hamiltonian of RGE}

When RGE is a Hamiltonian system, it is written as
\begin{eqnarray}
        \label{f1}
        \F{\D\B1}{\D t} &=& f_1 (\B1,\C1,\B2,\C2;\epsilon) 
        = \fracp{\HRG}{\C1}, \\
        \label{g1}
        \F{\D\C1}{\D t} &=& g_1 (\B1,\C1,\B2,\C2;\epsilon) 
        = -\fracp{\HRG}{\B1},\\
        \label{f2}
        \F{\D\B2}{\D t} &=& f_2 (\B1,\C1,\B2,\C2;\epsilon)
        = \fracp{\HRG}{\C2}, \\
        \label{g2}
        \F{\D\C2}{\D t} &=& g_2 (\B1,\C1,\B2,\C2;\epsilon)
        = -\fracp{\HRG}{\B2}.
\end{eqnarray}
By integrating equation (\ref{f1}) with respect to $\C1$,
 we have
\begin{equation}
        \label{step1}
        \HRG = \HRGN{(1)} + \MN{(1)}(\B1,\B2,\C2),
\end{equation}
\begin{equation} 
        \HRGN{(1)} = \int f_1 \D \C1,
\end{equation}
where $\MN{(1)}$ is an arbitrary function of $\B1,\B2$ and $\C2$.
%%%%
To determine $M_1$, we substitute equation (\ref{step1}) to
equation (\ref{g1}) and we obtain
\begin{equation} 
        \label{M1-diff}
        \fracp{\MN{(1)}}{\B1} = \left.\left( 
        -\fracp{\HRGN{(1)}}{\B1} - g_1 \right) \right|_{\C1=0}
\end{equation}
since the left-hand-side of equation (\ref{M1-diff}) does not depend on
$\C1$ when the RGE is a Hamiltonian system.
By integrating equation (\ref{M1-diff}) with respect to $\B1$, 
we get $\MN{(1)}$ as
\begin{equation} 
        \label{M1}
        \MN{(1)}(\B1,\B2,\C2) = \HRGN{(2)} + \MN{(2)}(\B2,\C2),
\end{equation}
where 
\begin{equation} 
        \HRGN{(2)} = \int \left.\left( 
        -\fracp{\HRGN{(1)}}{\B1} - g_1 \right) \right|_{\C1=0} \D \B1.
\end{equation}

Next, to determine $\MN{(2)}(\B2,\C2)$, we substitute equations 
(\ref{step1}) and (\ref{M1}) to equation (\ref{f2}), then
\begin{equation} 
        \label{M2-diff}
        \fracp{\MN{(2)}}{\C2} = \left.\left( 
        -\fracp{}{\C2}(\HRGN{(1)}+\HRGN{(2)}) + f_2 
        \right) \right|_{\B1=\C1=0}
\end{equation}
By integrating equation (\ref{M2-diff}) with respect to $\C2$, 
we get $\MN{(2)}$ as
\begin{equation} 
        \label{M2}
        \MN{(2)}(\B2,\C2) = \HRGN{(3)} + \MN{(3)}(\B2),
\end{equation}
where 
\begin{equation} 
        \HRGN{(3)} = \int \left.\left( 
        -\fracp{}{\C2}(\HRGN{(1)}+\HRGN{(2)}) + f_2 
        \right) \right|_{\B1=\C1=0} \hspace*{-1em}\D\C2.
\end{equation}

Finally, to determine $\MN{(3)}$, we substitute equations
(\ref{step1}), (\ref{M1}) and (\ref{M2}) to equation (\ref{f2}), 
then
\begin{equation} 
        \label{M3-diff}
        \fracp{\MN{(3)}}{\B2} = \left.\left( 
        -\fracp{}{\B2}(\HRGN{(1)}+\HRGN{(2)}+\HRGN{(3)}) - g_2 
        \right) \right|_{\B1=\C1=\C2=0}
\end{equation}
By integrating equation (\ref{M3-diff}) with respect to $\B2$, 
we get $\MN{(3)}$ as
\begin{equation} 
        \label{M3}
        \MN{(3)}(\B2) = \HRGN{(4)},
\end{equation}
where 
\begin{equation} 
        \HRGN{(4)} = \int \left.\left( 
        -\fracp{}{\C2}(\HRGN{(1)}+\HRGN{(2)}+\HRGN{(3)}) - g_2 
        \right) \right|_{\B1=\C1=\C2=0} \hspace*{-2em}\D\B2.
\end{equation}
Consequently, the Hamiltonian $\HRG$ is
\begin{equation}  
        \HRG = \HRGN{(1)} + \HRGN{(2)} + \HRGN{(3)} + \HRGN{(4)}.
\end{equation}
%%%%%%%%%%%%%%%%%%%%%%%%%%%%%%%%%%%%%%%%%%%%%%%%%%%%%%%%%

\subsection{RGE Hamiltonian for quartic perturbation}

We present explicit form  of Hamiltonian of RGE for the system
%%%%
\begin{equation}
 \fl\hspace*{2em}
        H = \F{1}{2}(p_1^2 + p_2^2 + q_1^2 + q_2^2 )
         + \epsilon (\al1 q_1^4 + \al2 q_1^3 q_2 + \al3 q_1^2 q_2^2
                + \al4 q_1 q_2^3 + \al5 q_2^4).
\end{equation}
%%%%
We know that RGE for this system has Hamiltonian $\HRG$ if the 
coefficients $\al1,\cdots,\al5$ satisfy the following
condition \cite{yamaguchi-98b}:
%%%%%%%%%%%%%%%
\begin{equation}
\eqalign{
   & 9\al2^2+4\al3^2-24\al1\al3-9\al2\al4=0, \\
   & 9\al4^2+4\al3^2-24\al3\al5-9\al2\al4=0, \\
   & (\al2+\al4)\al3-6(\al1\al4+\al2\al5)=0.
}
\end{equation}
%%%%
 Then the $\HRG$ is
%%%%
\begin{equation}
  \HRG = \epsilon \HRG_1 + \epsilon^2 \HRG_2,
\end{equation}
%%%%
where
\begin{eqnarray}
     \fl
    \HRG_1
        = \F{3}{8} \left\{
                \al1 (\B1^2+\C1^2)^2 + \al5 (\B2^2+\C2^2)^2 \right.\NO
                \left. + [ \al2 (\B1^2+\C1^2) + \al4 (\B2^2+\C2^2) ]
                        (\B1\B2+\C1\C2) \right\} \NO
        + \F{1}{8} \al3 (\B1^2+\C1^2) (\B2^2+\C2^2) 
        + \F{1}{4} \al3 (\B1\B2+\C1\C2)^2 ,
\end{eqnarray}
and 
\begin{equation}
\eqalign{
   \fl \HRG_2
      = -\F{5}{32} [ \al1^2 (\B1^2+\C1^2)^3 
                + \al5^2 (\B2^2+\C2^2)^3 ] \\
        \hspace*{-1.5cm} \lo- \F{5}{512} [ 
                        \al2^2 (\B1^2+\C1^2)^3 + \al4^2 (\B2^2+\C2^2)^3 ] \\   
        \hspace*{-1.5cm} \lo- \F{45}{512} [ \al2^2 (\B1^2+\C1^2)^2 
(\B2^2+\C2^2)
                + \al4^2 (\B1^2+\C1^2) (\B2^2+\C2^2)^2 ] \\
        \hspace*{-1.5cm} \lo- \F{5}{128} \al3^2 (\B1^2+\C1^2)
(\B2^2+\C2^2)
                (\B1^2+\B2^2+\C1^2+\C2^2) \\
        \hspace*{-1.5cm} \lo- \F{15}{64} [ \al1\al2 (\B1^2+\C1^2)^2 
                + \al4\al5 (\B2^2+\C2^2)^2 ] (\B1\B2+\C1\C2) \\
        \hspace*{-1.5cm} \lo- \F{15}{256} \al2\al4
(\B1\B2+\C1\C2+\B1\C2-\B2\C1) \\
                \times (\B1\B2+\C1\C2-\B1\C2+\B2\C1)(\B1^2+\B2^2+\C1^2+\C2^2) \\
        \hspace*{-1.5cm} \lo- \F{5}{32} [ \al1\al3 (\B1^4-\C1^4)
(\B2^2-\C2^2)
                + \al3\al5 (\B1^2-\C1^2) (\B2^4-\C2^4) ] \\
        \hspace*{-1.5cm} \lo- \F{5}{8} [ \al1\al3 (\B1^2+\C1^2) 
                + \al3\al5 (\B2^2+\C2^2) ] \B1\B2\C1\C2 \\
        \hspace*{-1.5cm} \lo- \F{5}{64} (\al1\al4+\al2\al5)
(\B1\B2+\C1\C2)^3 \\
        \hspace*{-1.5cm} \lo- \F{5}{128} \left\{ \al2\al3 
                [ (\B1^2+\C1^2)^2 + 3 (\B1^2\B2^2+\C1^2\C2^2) \right. \\
        \hspace*{-1.5cm} \hspace*{6em} \lo+ 2(\B1\C2+\B2\C1)^2 -2\B1\B2\C1\C2 ]  \\
        \hspace*{-1.5cm} \hspace*{1.5em} + \al3\al4
                 [ (\B2^2+\C2^2)^2 + 3 (\B1^2\B2^2+\C1^2\C2^2) \\
        \hspace*{-1.5cm} \hspace*{6em} \left. 
                 \lo+ 2(\B1\C2+\B2\C1)^2 -2\B1\B2\C1\C2 ] \right\} \
               (\B1\B2+\C1\C2).
}
\end{equation}

\end{document}